\documentstyle[12pt,epsf]{article}
\begin{document}

\newcommand{\be}{\begin{equation}}
\newcommand{\ee}{\end{equation}}
\newcommand{\ba}{\begin{eqnarray}}
\newcommand{\ea}{\end{eqnarray}}
\newcommand{\non}{\nonumber\\ }
\newcommand{\eq}[1]{(\ref{#1})}

\renewcommand{\thefootnote}{\fnsymbol{footnote}}
\font\csc=cmcsc10 scaled\magstep1
{\baselineskip=14pt
 \rightline{
 \vbox{\hbox{UT-800}
       \hbox{December 1997}
}}}

\vfill
\begin{center}
{\large\bf
BPS Condition of String Junction from M Theory
}

\vfill

{\csc Yutaka MATSUO}\footnote{
      e-mail address : matsuo@phys.s.u-tokyo.ac.jp},
{\csc Kazumi OKUYAMA}\footnote{
      e-mail address : okuyama@hep-th.phys.s.u-tokyo.ac.jp}\\
\vskip.1in

{\baselineskip=15pt
\vskip.1in
  Department of Physics,  Faculty of Science\\
  Tokyo University\\
  Bunkyo-ku, Hongo 7-3-1, Tokyo 113, Japan
\vskip.1in
}

\end{center}
\vfill

\begin{abstract}
{
We give a simple derivation of BPS condition of
string junction from M theory
}
\end{abstract}
\vfill

hep-th/9712070
\setcounter{footnote}{0}
\renewcommand{\thefootnote}{\arabic{footnote}}
\newpage
\vfill

Recently BPS condition of the $(p,q)$-string 
junction\footnote{Similar $(p,q)$-fivebrane junction 
is considered  in ref.\cite{r:aharony}} 
conjectured in \cite{r:Schwarz}
was derived in the linearized approximation 
by Dasgupta and Mukhi \cite{r:DM}.
Such a configuration seems physically relevant
in the description of the exceptional groups from open
string theory \cite{r:GZ}.
It might be also important in the understanding
of manifestly $SL(2,Z)$ invariant formulation in
type IIB superstring theory \cite{r:T}.
Subsequently further study was made on the string network
\cite{r:Sen} and BPS dynamics \cite{r:RY}.

BPS condition in \cite{r:DM} may be summarized as follows.
We are considering a junction of three
$(p_i,q_i)$ strings ($i=1,2,3$) (Fig.1)
stretching in $\vec{n}_i$ directions in $X^1$-$X^2$ plane 
($|n_i|=1$). 
\begin{figure}[h]
       \epsfxsize=8cm
       \centerline{\epsfbox{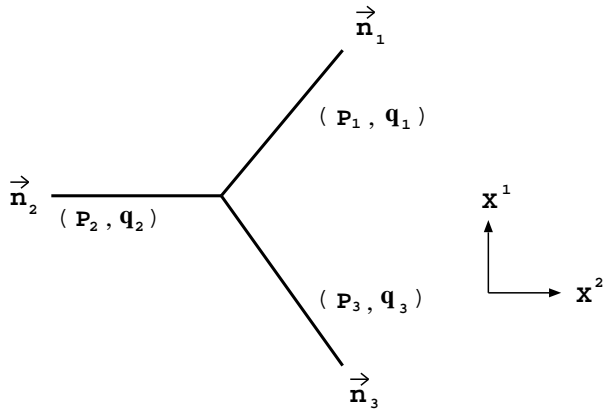}}
       \caption{String Junction}
\end{figure}
To keep supersymmetry, we need to impose
\begin{enumerate}
 \item the string charges should be
conserved \cite{r:Schwarz},
\be\label{e:charge}
\sum_{i=1}^3 p_i=
\sum_{i=1}^3 q_i= 0.
\ee
 \item the forces from each string tensions should be balanced
 \cite{r:DM},
\be\label{e:tension}
\sum_{i=1}^3 T_i \vec{n}_i=0, \quad
T_i=|p_i \tau +q_i|,
\quad \tau=\chi+i e^{-\phi}.
\ee
\end{enumerate}

In this short note, we will attempt to give
a simplified derivation of BPS condition from
M-theory viewpoint\cite{r:ASY}.
BPS configuration of M-theory (or more exactly 
supermembrane) was discussed by using $\kappa$-symmetry
\cite{r:BBP} when target space is Calabi-Yau manifold.
For the toroidal target space, it was examined in
\cite{r:EMM} in the light cone gauge 
and found that the supersymmetric configurations
are  essentially restricted to
double dimensional reduction of the string BPS configuration,
namely holomorphic map in two directions and trivial
winding in third direction.\footnote{ We note that
similar construction of BPS state was used by Witten\cite{r:Witten}
in the derivation of Seiberg-Witten curve from fivebrane configuration.}
We show that the requirement of holomorphy directly gives
the BPS condition discussed above.

Recall that the $(p,q)$ string is described in $M$ theory 
as a membrane winding $q$ (resp. $p$) times in $X^{11}$ (resp. $X^9$)
direction.  Namely if we parametrize the spatial part of 
the world volume of membrane
by $\sigma_1,\sigma_2$ where the winding direction described by
$\sigma_2$, the boudary condition is given as,
\begin{equation}
W(\sigma_1,\sigma_2+2\pi)  =   W(\sigma_1,\sigma_2)+ 
p\tau+q, \qquad W=X^{11}+i X^{9}.
\end{equation}
The string junction is then described as a ``pants'' diagram
(Fig.2 left). In the infinite leg limit, this diagram is
holomophically equivalent to a sphere with three punctures
(Fig.2 right)
\begin{figure}[h]
       \epsfxsize=6cm
       \centerline{\epsfbox{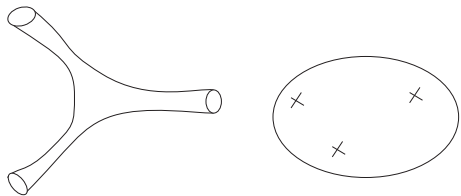}}
       \caption{Pants diagram}
\end{figure}
We take a holomorphic coordinate $\zeta$ on this sphere where
the three legs are attached to points $\zeta_i$ ($i=1,2,3$).
From the mapping in Fig.2, the asymptotic 
behaviour of the embedding into target space
near each singularity should look like,
\begin{eqnarray}
Z & \equiv & X^1+i X^2 \sim {\cal S}_i\tilde{\sigma}^{(i)}_1,\non
W & \sim & {\cal T}_i \tilde{\sigma}^{(i)}_2,
\end{eqnarray}
where
$
{\cal T}_i =  p_i \tau+q_i$,
$
\mbox{arg}({\cal S}_i)  =  \theta_i 
$ with
$$ \vec{n}_i=
\left(\begin{array}{c}
 \cos \theta_i\\ \sin \theta_i
      \end{array}\right),\quad
\log(\zeta-\zeta_i)=\tilde{\sigma}^{(i)}_1 +
i \tilde{\sigma}^{(i)}_2.$$
This embedding does not look like holomorphic 
as long as we take $W$ and $Z$ as holomorphic variables.
Our strategy is to replace it by an appropriate
complex structure in this four dimensional target space
such that the above embedding becomes holomorphic.
This condition actually gives over-constrained system of
equations and the BPS conditions \eq{e:charge}, \eq{e:tension} 
becomes conditions where they have
solutions.

We write two complex numbers $W,Z$ as four dimensional vector,
$\vec{Y}=(X^{11}, X^9, X^1, X^2)^t$ and let $J$ be a complex structure
described by four by four matrix satisfying,
\be\label{e:J}
J^2 = -I_4, \quad J^t \cdot J = I_4.
\ee
Cauchy-Riemann relation can be written as,
\be\label{e:CR}
\partial_{\sigma_1} \vec{Y} = \partial_{\sigma_2}(J\vec{Y}),
\ee
where we choose a particular holomorphic coordinate
of the original Riemann sphere $\zeta=\sigma_1+ i \sigma_2$.
At the asymptotic region, we may replace
them by $\tilde{\sigma}^{(i)}_{\{1,2\}}$.
\eq{e:CR} then gives,
\be
\left(
\begin{array}{c}
0 \\ 0 \\ S^{(i)}_1 \\ S^{(i)}_2
\end{array}\right) = J\cdot
\left(
\begin{array}{c}
T^{(i)}_1 \\ T^{(i)}_2\\
0\\0\end{array}\right),\quad
{\cal S}^{(i)}=S^{(i)}_1 + i S^{(i)}_2,\quad
{\cal T}^{(i)}=T^{(i)}_1 + i T^{(i)}_2.
\ee
One may write $J$ as 
\be
J=\left(\begin{array}{cc}
0 & m_1 \\
m_2 & 0 \end{array}\right).
\ee
The conditions (\ref{e:J}) reduce to following conditions on $m$,
$$
m_1 \cdot m_2 = -I_2, \quad m_i\cdot m_i^t= I_2.
$$
Since matrix $m$ now belongs to $O(2)$, 
one can immediately recognize that Cauchy-Riemann
relation gives  
\be
S^{(i)}_1+ iS^{(i)}_2 = e^{i\varphi} (T_1^{(i)}\pm iT_2^{(i)}).
\ee
Namely, up to rotation and parity in $X^1$-$X^2$ plane, 
the spacial orientation of each string is determined from
its charge $(p,q)$.  This condition seems stronger than the
requirement \eq{e:tension}.  However, as observed in \cite{r:Sen}
they are equivalent.   Finally, the charge conservation \eq{e:charge}
comes from the integration,
\be
\sum_i \int_{C_i} d(\vec{Y}+iJ\vec{Y})=0.
\ee
where $C_i$ is a circle which surrounds the point $\zeta=\zeta_i$.

Although we used only the asymptotic behaviour of the embedding,
it is obvious that such embedding exits globally
from the very nature of holomorphy.
This completes our derivation of BPS condition.
At this point, it is obvious to extend our analysis to
$n$-string junction or ``string network'' discussed by Sen
\cite{r:Sen} (the holomophic map of a Riemann surface
with $n$ punctures).  

After we finished our computation, we saw a paper \cite{r:related}
which discuss the string junction in the similar line.


\end{document}